\documentstyle[prl,twocolumn,aps]{revtex}
\begin{document}
\title{Long-range correlations in the wave functions of chaotic systems}
\author{Vladimir I. Fal'ko $^{1,3,4}$ and K.B. Efetov$^{2,5}$}
\address{$^1$ Dept. of Theoretical Physics, Oxford University, 1 Keble
Road, OX1 3NP, Oxford, UK\\ 
$^2$ Max-Planck-Institut f\"ur Physik komplexe Systeme,
Heisenbergstr.1, 70569 Stuttgart, Germany \\ 
$^3$ School of Physics, Lancaster University, LA1 4YW Lancaster, UK\\ 
$^4$ Institute of Solid State Physics RAS, Chernogolovka, 142432
Russia\\ 
$^5$ L.D. Landau Institute for Theoretical Physics, Kosygina 1,
Moscow, Russia} 
\date{\today{}}
\maketitle

\begin{abstract}
We study correlations of the amplitudes of wave functions of a chaotic
system at large distances. For this purpose, a joint distribution function
of the amplitudes at two distant points in a sample is calculated
analytically using the supersymmetry technique. The result shows that,
although in the limit of the orthogonal and unitary symmetry classes the
correlations vanish, they are finite through the entire crossover regime and
may be reduced only by localization effects.
\end{abstract}


During the past decade, a substantial progress has been achieved in the
studies of quantum properties of classically chaotic systems \cite
{GiannoniGutzwillerHaake}. At present, we have quite a complete information
about energy spectra of non-integrable systems in the limit of a ''hard''
chaos as well as in the crossover regimes to integrability or localization.
However, the theoretical background for understanding the statistics and
structure of wave functions of chaotic systems is still a growing up field
stimulated by the recent experiments on electron transport in quantum dots
under the Coulomb blockade conditions \cite{Chang}, as well as by the
studies of the microwave irradiation in disordered and chaotic
cavities 
\cite{Sridhar}. According to the resonant tunneling models of the transport in
the Coulomb blockade devices \cite{Stone,Prigodin}, the height $g$ of the
conductance peaks measured in the experiments on the quantum dots can be
related to the local amplitudes $\varphi _\alpha ({\bf r}_{l,r})$ of the
wave function of the resonant level $\epsilon _\alpha $ as

\begin{eqnarray}
\label{e1}g=\frac{e^2/h}{4\pi T}\frac{\Gamma _l\Gamma _r}{\Gamma
_l+\Gamma _r},\:   
{\rm where}\: {\rm broadening} \: \Gamma _{l,r}\propto\left| 
\varphi _\alpha \left(
{\bf r}_{l,r}\right) \right| ^2 \nonumber
\end{eqnarray}
is due to the electron escape to the right- and left-hand-side electrodes 
\cite{Remark}. Hence, the statistics of the resonant tunneling conductances
in this kind of measurements is determined by the statistics of the electron
wave functions in the dot. Moreover, the statistics and spatial correlation
properties of microwaves in random media have become the subject of direct
measurements performed already in various parametric regimes \cite{Sridhar}.

Below, we demonstrate the existence of long-range correlations in the wave
functions of a chaotic system subjected to a weak magnetic field, that is,
in the crossover regime between the orthogonal (with time-reversal symmetry)
and unitary (broken time reversal symmetry) ensembles. This conclusion is
based on the analysis of the joint probability distribution function 
$f_2(p_1,p_2)$ which describes the statistics of local densities of an
eigenstate in a chaotic cavity taken at two distant points in the sample and
its comparison with the statistics of a single-point distribution 
$f_1(p_1)$. The joint distribution functions $f_N(p_1,...,p_N)$ of
local densities of 
a wave function are defined for an arbitrary $N$ as follows

\begin{equation}
\label{e2}f_N=\Delta \left\langle \sum_\alpha \delta\left( \epsilon
-\epsilon _\alpha \right) \prod_{n=1}^N \delta \left( p_n-\left| \varphi
_\alpha \left( {\bf r}_n \right) \right| ^2V\right) \right\rangle , 
\end{equation}
where the angular brackets stand for averaging over irregularities in the
system or over energies $\epsilon$ \cite{Andreev} in a chaotic billiard, $V$
is the volume of the system and $\Delta$ is the mean level spacing. The sum
in Eq. (\ref{e2}) is taken over all eigenstates, $\varphi_{\alpha}$
and $\epsilon_{\alpha}$ being eigenfunctions and eigenenergies. The joint
probability distribution function $f_2$ naturally appears in the description
of fluctuations of the resonant tunneling conductance peaks, $g$, in quantum
dots or can be directly studied in the microwave experiments \cite{Sridhar}.

Direct computation of the local density distribution function $f_1$ for
disordered and chaotic systems in the limiting cases of the orthogonal and
unitary ensembles \cite{Prigodin,Efetov3,FalkoEfetovPR} leads to the well
known Porter-Thomas distribution\cite{Brody} which is merely a Gaussian
distribution of wave function amplitudes. The Gaussian distribution has been
recently confirmed by the numerical studies of high-lying chaotic
eigenstates \cite{Robnik}. It can also be derived semiclassically assuming
that classical orbits cover uniformly the energy surface in the phase space 
\cite{Berry0}. The detailed analysis of the joint distribution function $f_2$
performed in Refs. \cite{Prigodin2} for these two universality classes has
shown that intrinsic correlations in the wave functions of a chaotic system
decay already on several wavelengths and the function $f_2$ can be written
at large distances as

\begin{equation}
\label{e4}f_2\left( p_1,p_2\right) \rightarrow f_1\left( p_1\right)
f_1\left( p_2\right) 
\end{equation}
where the function $f_1\left( p\right)$ describes the Gaussian distribution
of local amplitudes $\varphi ({\bf r}_n)$, 
$p_n\equiv V\left|\varphi ({\bf r}_n)\right|^2$. 
In particular, the latter result has been used in
Refs. \cite{Stone,Prigodin} upon deriving the distribution function of
conductance 
peaks in the orthogonal and unitary symmetry ensembles which were later
verified experimentally, Refs. \cite{Chang}. Moreover, one can provide a
full analytical description of a decay of correlations of local amplitudes
of chaotic wave functions in each of the symmetry classes 
\cite{Prigodin2,Srednicki}. The law describing the decay of the
correlations can 
be derived either using the supersymmetric $\sigma$-model technique or with
Berry's conjecture \cite{Berry0} about the uniform distribution of the
classical orbits in the phase space \cite{Srednicki}, and the result for the
orthogonal case is in a good agreement with that of the microwave
experiments \cite{Sridhar}.

The suppression of correlations within a chaotic wave function at large
distances (Eq. (\ref{e4})) seems to be so natural for systems in which the
semiclassical version of quantum mechanics deals with large random phases
that it was not called in question also in the regime of a crossover between
the orthogonal and unitary symmetry classes, where the explicit calculations
have not been made so far. For example, in Ref. \cite{Alhassid} it was
argued that the presence of a small magnetic field cannot considerably
change the law describing the decay of the correlations, and this assumption
was implicit in other works.

In this Letter we show that due to the Aharonov-Bohm effect, 
the intrinsic correlations implicit into a chaotic
wave function do not vanish at large distances in the entire 
crossover regime, although they do in the orthogonal and unitary
limits (corresponding to zero 
and sufficiently high magnetic fields, respectively). In weak magnetic
fields, the correlations survive at distances 
$r=\left| {\bf r}_1-{\bf r}_2\right|$ 
much longer than the mean free path $l$ (that is, 
$f_2(p_1,p_2)\ne f_1(p_1)f_1(p_2)$).

Calculations presented below are performed using the supersymmetry technique 
\cite{Efetov}. In particular, the use of this method enables one to
calculate the one-point distribution function $f_1(p)$ for an arbitrary
magnetic field \cite{FalkoEfetovPR} (a somewhat different although related
quantity was calculated in Ref. \cite{IidaSommers}). The function $f_2$ can
be computed in a similar way. Although the calculations are not more
difficult than those in Ref. \cite{FalkoEfetovPR} they lead to the result
about the long range correlations which could not be extracted from the
function $f_1$ alone.

To carry out the computations within the supersymmetry scheme, one should
write the quantities to be studied in terms of the Green functions defined,
as usual, as

\begin{equation}
\label{e5}G_{\epsilon}^{R,A}({\bf r}_1,{\bf r}_2)=\sum_{\alpha}
\varphi_{\alpha}({\bf r}_1)\varphi^{*}_{\alpha}({\bf r}_2) / \left[\epsilon
-\epsilon_{\alpha}\pm i\gamma /2 \right] . 
\end{equation}
It is not difficult to show that the distribution function $f_2$, Eq. (\ref
{e2}), can be represented as follows:

\begin{eqnarray}\label{e6}
f_2(p_1,p_2)=\frac{\Delta}{\pi}
\lim_{\matrix{_{\beta\to 1}\cr _{\gamma\to 0}\cr}} 
\frac{\partial}{\partial\beta }
\langle\beta\int_0^1 dt\int d{\bf r} 
\Im\left[ G_{\epsilon}^A({\bf r,r})\right] \nonumber \\
\delta \left[ p_1+iY_AG_{\epsilon}^A({\bf r}_1,{\bf r}_1)\right] 
\delta \left[ p_2-iY_RG_{\epsilon}^R({\bf r}_2,{\bf r}_2)\right] 
\rangle , 
\end{eqnarray}
where $Y_A=\gamma \beta Vt$ and $Y_R=\gamma \beta V(1-t)$. Next, we express
the Green functions entering Eq. (\ref{e6}) in terms of integrals over
supervectors $\psi $ with the Lagrangian $L$ 
\begin{equation}
\label{e7}L\left[ \psi \right] =i\int\bar\psi ({\bf r})\left[ \epsilon -\hat 
H_0-U({\bf r})-i\gamma\Lambda /2\right] \psi ({\bf r}) d{\bf r} 
\end{equation}
where $\hat H_0$ is the free particle Hamiltonian and $U({\bf r})$ describes
the impurity potential. The diagonal matrix $\Lambda$ has components 
$\Lambda^{11}=-\Lambda^{22}=1$ and $\psi$ is an $8$ -component supervector
containing both commuting $s$ and anticommuting $\chi$ elements. The
reduction of Eq. (\ref{e6}) to the integral over the supervectors $\psi$ can
be done by expanding $\delta$-functions in the Green functions 
$G_{\epsilon}^R$ and $G_{\epsilon}^A$ 
and using standard formulae of Gaussian
integration. Then one averages over the irregularities and, finally, uses
the Hubbard-Stratonovich transformation to decouple the ''interaction'' term 
$(\bar\psi\psi )^2$ obtained after this averaging by an integration over the
supermatrix field $Q$. This gives us a possibility to integrate over the
supervectors $\psi$ and to reduce the following computation to that of an
integral over the supermatrices $Q$. Integrating over the supervector $\psi$
we take into account only pairing at coinciding points. The approximation is
justified provided the distance between the points ${\bf r}_1$ and 
${\bf r}_2$ is large enough (much larger than the wavelength). 
This means that we
consider the asymptotic form of the distribution function of local
amplitudes of the wave function taken at essentially distant
points. All manipulations are standard \cite{Efetov,FalkoEfetovPR},
and 
we arrive at

\begin{eqnarray}\label{e9}
f_2(p_1,p_2)=\lim_{\matrix{_{\beta\to 1}\cr _{\gamma\to 0}\cr}} 
\frac{d}{d\beta}\langle\frac{\beta}{2}\int
\frac{d\zeta_1d\zeta_2}{(2\pi )^2}
\int_0^1dt\left( Q_{11}^{11}-Q_{11}^{22}\right)
\nonumber \\
\delta\left( p_1-\frac{t\gamma\beta}{2\Delta}\bar z_1Qz_1\right) 
\delta\left( p_2+\frac{\left( 1-t\right)\gamma\beta }{2\Delta }
\bar z_2Qz_2\right)\rangle_{_Q}, 
\end{eqnarray}
where $\left\langle ...\right\rangle _Q$ stands for the integration over
supermatrices $Q$ with the free energy $F\left[ Q\right]$, and 
$\bar z_2=(0,0,0,0,0,0,e^{i\zeta _2},e^{-i\zeta _2})$, 
$\bar z_1=(0,0,e^{i\zeta_1},e^{-i\zeta _1},0,0,0,0)$. 
In the presence of a magnetic field, $F[Q]$ is given by \cite{Efetov1}.

\begin{eqnarray}
\label{e10}F=-{\rm Str}\left( \frac{\pi \gamma }{4\Delta }\Lambda
Q+\left( \frac {X}{4}\right) ^2\left[ Q,\tau _3\right] ^2\right) 
,\\ 
X^2=2\pi \frac D\Delta \int \frac{d{\bf r}}V\left( 
\frac{2\pi {\bf A}\left( 
{\bf r}\right) }{\phi _0}\right) ^2=\alpha _g\frac{\phi ^2}{\phi _0^2}
\frac{E_c}{\Delta} 
\nonumber
\end{eqnarray}
where $\phi_0=hc/e$ is the flux quantum, $\alpha _g$ is the sample geometry
dependent factor, $D$ is the classical diffusion coefficient, $E_c$ is the
Thouless energy. Eqs. (\ref{e9},\ref{e10}) are written under an assumption
that the supermatrix $\sigma$-model describing fluctuations of the
supermatrix $Q$ is zero-dimensional (0D), which is correct for not very
large $p_{1,2}$. The distribution function $f_2$ can also be calculated for
arbitrary $p_{1,2}$ in a way that it has been done for the function $f_1$ in
Ref. \cite{FalkoEfetovEL}. This would require going beyond the 0D 
$\sigma$-model and is out of the scope of this publication.

Calculation of the integral over $Q$ in Eq. (\ref{e9}) can be carried out
using the parametrization suggested in Ref. \cite{Efetov1} which separates
''cooperon'' and ''diffuson'' variables. The quantum limit $\gamma\to 0$
considerably simplifies the integration over the diffuson variables since
the main contribution comes from that part of the non-compact sector of the
order parameter $Q$ where 
$\left| \bar z_nQz_n\right|\sim 1/\gamma\to\infty$.
In this limit, the diffuson degrees of freedom can be integrated out and
one ends up with a version of some kind of a reduced $\sigma$-model adapted
for describing the properties of a single quantum state \cite{FalkoEfetovEL}
and containing only the integration over the cooperon variables,

\begin{eqnarray}
\label{e11}f_N(p_1,...,p_N)=\int_0^1d\lambda _c\int_1^\infty 
d\lambda_{1c}\exp 
\left( -F_\phi \right) \\
\times 
\frac{\lambda _c^2}{\left( \lambda _{1c}^2-\lambda _c^2\right) ^2}
^{}\int \int P\prod_{n=1}^N {\frac{{d\zeta _n}}{{2\pi }}}{\frac{{\exp
(-p_n/A_n)}}{{A_n}}}dR_c. \nonumber
\end{eqnarray}
In Eq. (\ref{e11}), $N=1,2$, functions $P$ and $A_n$ are defined as

\begin{eqnarray}
P=\lambda_{1c}+2\left( \eta _c\eta _c^{*}-\kappa _c\kappa _c^{*}\right)
\left( \lambda _{1c}-\lambda _c\right) , 
A_n=1+\Re\frac {Le^{i\zeta_n }}{P} ,\nonumber \\
L=\sqrt{\lambda_{1c}^2-1}\left( 1+2\eta_c\eta_c^{*}\right) \left(
1-2\kappa_c\kappa_c^{*}\right) +4\sqrt{1-\lambda_c^2}\eta_c\kappa_c,
\nonumber
\end{eqnarray}
and $dR_c=d\eta_cd\eta_c^{*}d\kappa_c^{*}d\kappa_c$, with $\eta_c$, 
$\eta_c^{*}$, $\kappa_c$, $\kappa_c^{*}$ being Grassmann anticommuting
variables.

Skipping technical details of the algebraic manipulations with Eq. 
(\ref{e11}), we are able to represent the joint distribution function 
$f_N$, $N=1,2$ in the form

\begin{equation}
\label{e12}f_N=\int_1^\infty B(X,x)\prod_{n=1}^N M(p_n,x)dx 
\end{equation}
where 
\begin{eqnarray}
B(X,x)=X^2\left[ (xX^2-1)\Phi_2(X)+\Phi_1(X)\right] 
e^{-X^2(x-1)}, \nonumber \\
M(p,x)=\sqrt{x}\exp (-px)I_0\left( p\sqrt{x^2-x} \right) \nonumber
\end{eqnarray}
Here, ${\rm I}_0(z)$ is the modified Bessel function, and 
$$
\Phi_1(X)=\frac{e^{-X^2}}{X}\int_0^Xe^{y^2}dy,\quad \Phi_2(X)=
\frac{1-\Phi_1(X)}{X^2}. 
$$

In order to establish the presence or absence of correlations between
fluctuations of the values 
$p_{1,2}\equiv\left|\varphi ({\bf r}_{1,2})\right|$ 
at distant points in the sample in the crossover regime,
the joint probability distribution function $f_2$ and its moments have to be
compared to the distribution function $f_1$. Explicit integration over $x$
in Eqs. (\ref{e12}) shows that in each of two limits $X\to 0$ and $X\to
\infty $, $f_2(p_1,p_2)$ takes a separable form and satisfies Eq. 
(\ref{e4}). 
That is, $f_2(p_1,p_2)=\Pi _{i=1,2}\frac{e^{-p_i/2}}{\sqrt{2\pi p_i}}$ in
the orthogonal case and $f_2(p_1,p_2)=\exp (-p_1-p_2)$ in the unitary one.
However, the distribution function is not separable at any finite magnetic
field, which means that even in the limit of a large distance between the
points ${\bf r}_1$ and ${\bf r}_2$ the correlations do not vanish.

To illustrate {\it the existence of long-range correlation implicit in each
individual wave function in a quantum billiard subjected to a weak magnetic
field\/}, we calculate the correlation function $K_{2,2}(X)$ defined
as \cite{Remark2}

\begin{eqnarray}
K_{2,2}\left( X\right) =\int_0^\infty dp_1dp_2 p_1^2p_2^2 \left[
f_2\left( p_1,p_2\right) -f_1\left( p_1\right) f_1\left( p_2\right) 
\right] \nonumber
\end{eqnarray}
Although an analytical expression for $K_{2,2}$ can be found from
Eqs. (\ref{e12}), 
we give here only its graphic representation. Fig. 1 illustrates how
the correlations evolve as function of a normalized flux through the sample
area. As one can see from Fig. 1, the correlations are present all over the
crossover regime: from the smallest to highest values of the normalized flux 
$\phi $. These correlation are never strong (of the order of one percent),
since the maximal value of the correlator $\max (K_2)\approx 0.05$ has to be
compared to the square of the second moment $P_2=\int dpp^2f_1(p)$ which
varies from the value of $3$ to $2$ in between of the orthogonal and unitary
symmetry classes.

The distant correlations in the wave functions can be also traced in the
joint probability to find simultaneously two zeros of the wave
function 
$\varphi ({\bf r}_{1,2})=0$. 
At small fluxes corresponding to $X\ll 1$, 
$[f_1(0)]^2\approx \frac{4\pi}{9}X^{-2}$ whereas 
$f_2(0,0)\approx\frac{5}{3}X^{-2}$. 
Since the crossover between the orthogonal and unitary ensembles
occurs at relatively small magnetic fluxes through the sample area,
$\phi\sim\phi_0(\Delta /E_c)^{1/2}$, 
the non-separability of $f_2(p_1,p_2)$
as a whole is most pronounced when $X\sim 1$. In Fig. 2, we show the
difference between $f_2$ and $f_1\times f_1$ at $X=1$.

Starting from the pioneering work by Berry \cite{Berry0}, it was generally
believed that in classically ergodic systems the local wave function density
can be imagined as a result of superposition of an infinite number of plane
waves with random phases and equal momenta. The randomness of the phases
suggests the Gaussian randomness of the amplitude $\varphi_{\alpha}$ and
vanishing correlations at large distances. Our result means that a weak
magnetic flux through the sample area introduces some correlations in 
$\varphi_{\alpha}({\bf r})$ 
related to the fact that the Aharonov-Bohm phases
taken by an electron moving in a quantum billiard cannot be arbitrarily
large. The lengths of geometrical paths attributed to an electron classical
trajectory in a semiclassical picture of quantum mechanics are limited,
since the semiclassics breaks down at the time scale longer than the
Heisenberg time, $t_H\sim h/\Delta$. Hence, the encircled magnetic field
fluxes are also limited, so that the classical ergodicity does not always
lead to the complete randomness of the wave function phases. 
Another, better
known illustration of this fact can be given by a disordered system with a
ring geometry. In the latter case, the flux through the ring can exist even
if the magnetic field in the wire cross-sectional area is negligibly small,
which generates regular periodic oscillations of the wave function phases.

The amplitude correlations discussed above persist all over the chaotic
quantum billiard. One may ask a question about what happens if the sample
size is much larger than the mean free path in a disordered system. In fact,
the result of Eq. (\ref{e12}) is valid as far as the 0D $\sigma $-model may
be used, that is, unless localization effects become important. Hence, in 1D
and 2D weakly disordered samples the distance between the observation points
must be smaller than the localization length, whereas in a 3D metal the
distance between them is not limited. The correlations can be considerably
reduced also for large values of amplitudes originated from prelocalized
states. In this case, one has to go beyond the 0D $\sigma$-model 
\cite{DKh,FalkoEfetovEL}.

In conclusion, the joint distribution function of amplitudes of chaotic wave
functions is derived for an arbitrary magnetic flux. Its form manifests the
long-range spatial correlations existing in the entire crossover regime from
orthogonal to unitary symmetry classes. This indicates that due to the
Aharonov-Bohm effect the phases of the wave functions are correlated even if
the corresponding classical motion is ergodic, which has not been
anticipated in previous semiclassical theories.

Authors are grateful to V.Prigodin for discussions.  One of us (VF) thanks
EPSRC for a financial support and MPI-PKS for hospitality.

\begin{figure}
\caption{ The correlation function $K_{2,2}(X)$ which can be used as a
measure of spatial correlation implicit to a chaotic wave function in
the crossover regime from the orthogonal to unitary symmetry
classes. \label{Fig1} }
\caption{ The difference between $f_2$ and $f_1\times f_1$
normalized by $f_1\times f_1$ for $X=1$.
\label{Fig2} }
\end{figure}


\begin{references}
\bibitem{GiannoniGutzwillerHaake}  
M-J. Giannoni, J. Voros and Zinn-Justin
(eds.) {\it Chaos and Quantum Systems} (Amsterdam: North-Holland, 1991);
M.C. Gutzwiller {\it Chaos in Classical and Qunatum Mechanics} (New York:
Springer, 1990); F. Haake {\it Quantum Signatures of Chaos} (Berlin:
Springer, 1992) 

\bibitem{Chang}  
A.M. Chang {\it et al}, preprint (1995); J.A. Falk {\it et
al}, preprint (1995)

\bibitem{Sridhar}  
A. Kudrolli, V. Kidambi, S. Sridhar, Phys. Rev. Lett. 
{\bf 75}, 822 (1995); V.N. Prigodin {\it et al, ibid.} 
{\bf 75}, 2392 (1995)

\bibitem{Stone}  
R.G. Jalabert, A.D. Stone, and Y. Alhassid, Phys. Rev.
Lett. {\bf 68}, 3468 (1992)

\bibitem{Prigodin}  
V.N.Prigodin, K.B.Efetov, S.Iida, {\it ibid.} {\bf 71},
1230 (1993)

\bibitem{Remark}  
This formula for the resonance conductance corresponds to
the limit of $T\gg \Gamma $. Other regimes were studied in 
\cite{Prigodin}.

\bibitem{Andreev}  
O. Agam, B.L. Altshuler, A.V. Andreev, Phys. Rev. Lett. 
{\bf 75}, 4389 (1995); A.V. Andreev {\it et al}, preprint (1996)

\bibitem{Efetov3}  
K.B. Efetov and V.N. Prigodin, Phys. Rev. Lett. {\bf 70}, 1315 (1993); 
Mod. Phys. Lett. B {\bf 7}, 981 (1993);
A.D. Mirlin and Y.V. Fyodorov, J. Phys. A {\bf 26}, L551 (1993)

\bibitem{FalkoEfetovPR}  
V.I.Fal'ko, K.B.Efetov, Phys. Rev. B {\bf 50}, 11267 (1994)

\bibitem{Brody}  
T.A. Brody et al, Rev. Mod. Phys. {\bf 53}, 385 (1981)

\bibitem{Robnik}  
B. Li and M. Robnik, J. Phys. A {\bf 27}, 5509 (1994)

\bibitem{Berry0}  
M.V. Berry, J. Phys. A {\bf 10}, 2083 (1977)

\bibitem{Prigodin2}  
V.N. Prigodin, Phys. Rev. Lett. {\bf 74}, 1566 (1995);
E. Muciollo {\it et al, ibid.} {\bf 75}, 1360 (1995)

\bibitem{Srednicki}  
M. Srednicki, cond-mat/9512115

\bibitem{Alhassid}  
Y.Alhassid, C.Lewenkopf, Phys.Rev.Lett.{\bf 75},
3922(1995)

\bibitem{Efetov}  
K.B. Efetov, Adv. Phys. {\bf 32}, 53 (1983)

\bibitem{IidaSommers}  
H.-J. Sommers and S. Iida, Phys. Rev. E {\bf 49},
2513 (1994); K. Zyczkowsky and G. Lenz, Z. Phys. B {\bf 82}, 299 (1991)

\bibitem{Efetov1}  
A. Altland {\it et al}, Journ. Phys. A {\bf 26}, 3545
(1993); K.B. Efetov and S. Iida, Phys. Rev. B {\bf 47}, 15794 (1993)

\bibitem{DKh}  
B.A. Muzykantskii and D.E. Khmelnitskii, Phys. Rev. B {\bf 51}, 
5480 (1995)

\bibitem{FalkoEfetovEL}  
V.I. Fal'ko and K.B. Efetov,
Europhys. Lett. {\bf 32}, 627 (1995); Phys. Rev. B {\bf 52}, 17413 (1995)

\bibitem{Remark2}  
Because of the normalization of the wave functions 
$\varphi_{\alpha}$, $K_{2,2}$ is the lowest-order nontrivial correlation
function.
\end{references}
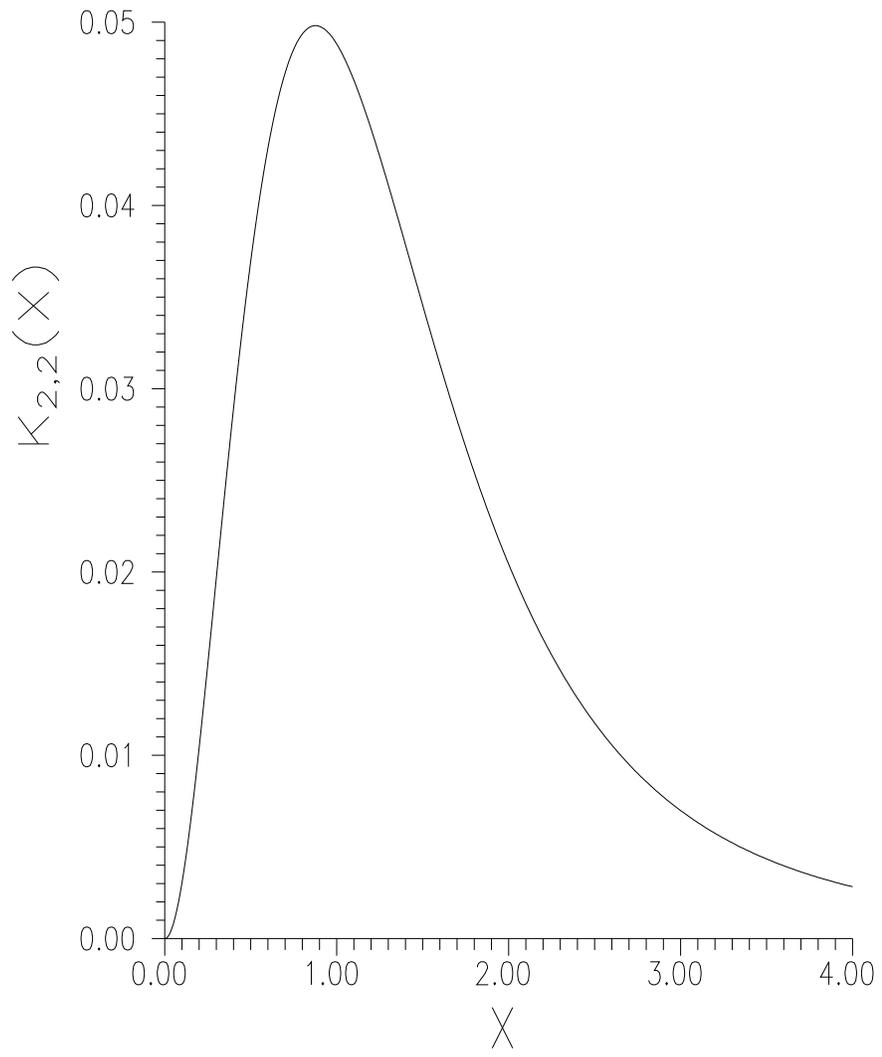
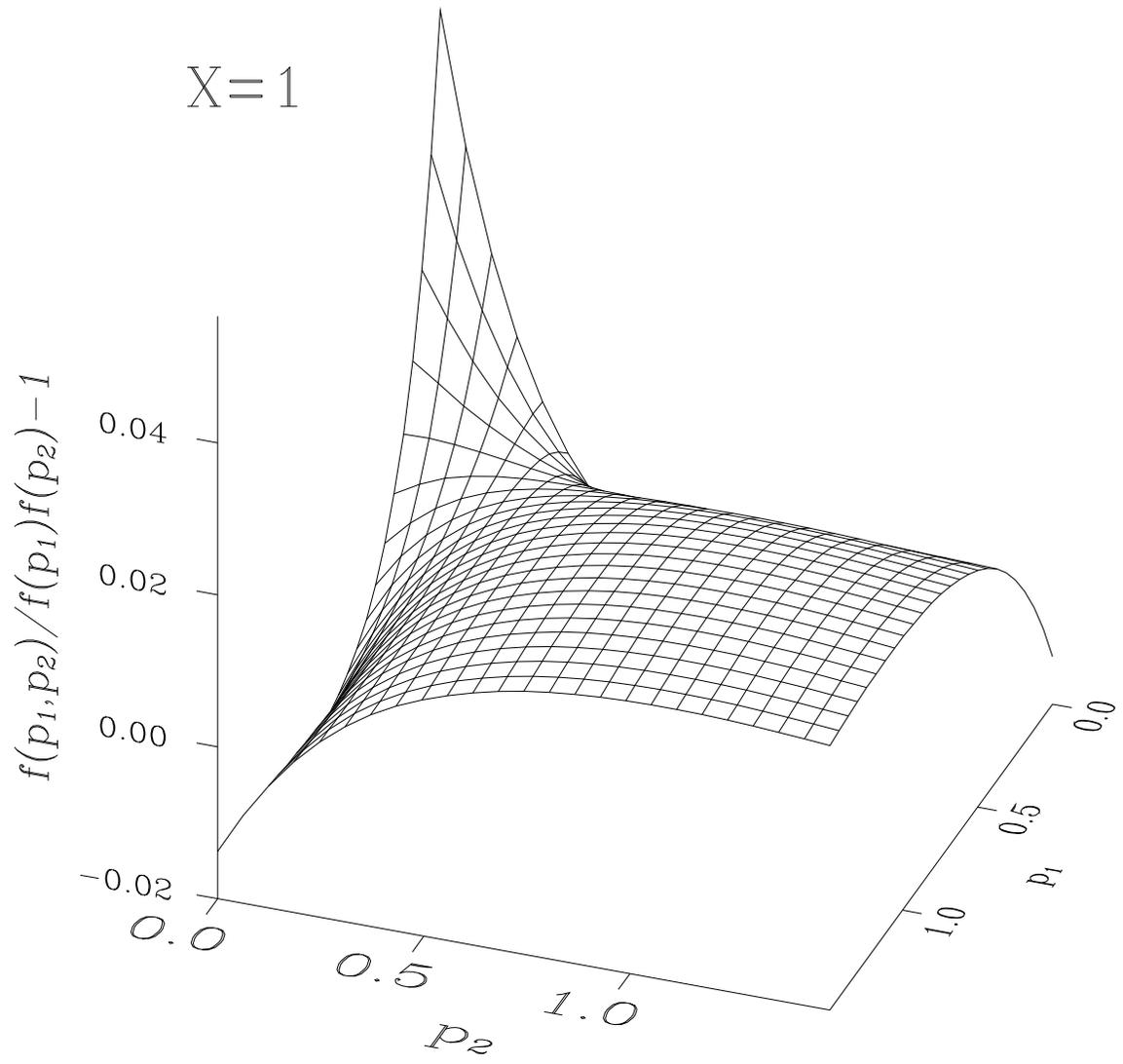
\end{document}